\documentclass[12pt,preprint]{aastex}
\usepackage{amsmath}
\usepackage{ulem}

\shorttitle{Hard X-ray excess of Mrk 421}
\shortauthors{Liang Chen}

\slugcomment{2017, ApJ, 842, 129}

\begin{document}
\title{On the origin of the hard X-Ray excess of high-synchrotron-peaked BL Lac object Mrk 421}

\author{Liang Chen{$^{1,2}$}}

\altaffiltext{1}{Key Laboratory for Research in Galaxies and
Cosmology, Shanghai Astronomical Observatory, Chinese Academy of
Sciences, 80 Nandan Road, Shanghai 200030, China; E-mail: chenliang@shao.ac.cn}
\altaffiltext{2}{Key Laboratory for the Structure and Evolution of Celestial Objects, Chinese Academy of Sciences, Kunming 650011, China}


\begin{abstract}

For the first time, Kataoka \& Stawarz reported a clear detection of a hard X-ray excess, above $\gtrsim$20 keV, in the high-synchrotron-peaked BL Lac object Mrk 421. We find that this feature may not be produced by the low-energy part of the same electron population that produced the {\it Fermi}/LAT $\gamma$-ray. Because of that it is required that the power-law electron energy go down to $\gamma_{\rm min}\approx19$, which predicts a very strong radio emission (radio flux larger than the observed) even considering the synchrotron self-absorption effect. We investigate the possibility of this excess being produced from the spine/layer jet structure, which has been clearly detected in Mrk 421. We find that (1) similar to one-zone modeling, the spine emissions provide good modeling of the broadband spectral energy distribution, except for the hard X-ray excess; and (2) the hard X-ray excess can be well represented by the synchrotron photons (from the layer) being inverse Compton scattered by the spine electrons.

\end{abstract}


\section{Introduction}
\label{sec:introduction}

Due to the Doppler beaming effect, relativistic jets dominate the broadband emissions (from radio to $\gamma$-ray) of a blazar, in which the viewing angle between the jet direction and line of sight is very small \citep[][]{1978bllo.conf..328B, 1995PASP..107..803U}. The broadband spectral energy distributions (SEDs) of blazars usually show two significant bumps. One of these peaks at infrared (IR) to X-ray bands, which is believed to be the synchrotron emissions of energetic electrons within the jet. The second bump peaks at the $\gamma$-ray band, which may be the inverse Compton (IC) emissions of the same distribution of electrons emitting the synchrotron bump \citep[e.g.,][see also the hadronic model: \citet{1993A&A...269...67M, 2000NewA....5..377A, 2003ApJ...586...79A, 2003APh....18..593M}]{1992ApJ...397L...5M, 1996ApJ...461..657B, 1994ApJ...421..153S, 1993ApJ...416..458D}.

The distribution of the synchrotron peak frequency, $\nu_{\rm sy}^{\rm p}$, forms a continuous sequence, although the objects usually classified as high-synchrotron-peaked (HSP; $\nu_{\rm sy}^{\rm p}\geq10^{15}$ Hz), intermediate-synchrotron-peaked (ISP; $10^{15}$ Hz $\geq\nu_{\rm sy}^{\rm p}\geq10^{14}$ Hz), and low-synchrotron-peaked (LSP; $\nu_{\rm sy}^{\rm p}\leq10^{14}$ Hz) blazars, depending on the location of $\nu_{\rm sy}^{\rm p}$ \citep{2010ApJ...715..429A}. The synchrotron and IC components usually cross at roughly the X-ray band. Therefore, for HSP blazars, the X-ray is usually dominated by softer synchrotron emissions. For most LSP blazars, the X-ray is almost dominated by harder IC emissions. For ISP blazars (and some LSP blazars), the X-ray is a mixture of synchrotron and IC emissions; therefore, the X-ray spectra sometimes show a concave shape in the $\log\nu-\log\nu f_{\nu}$ frame.

The SED modeling is a powerful tool for studying jet physics \citep[see, e.g.,][]{2009MNRAS.397..985G, 2010MNRAS.402..497G, 2014Natur.515..376G}. The lowest-energy electrons are the most plentiful and sensitive to probe the total content of particles in the jet, including the estimated jet power, the particle acceleration mechanism, etc. However, it is very difficult to use radio observation to constrain this minimal electron energy, even though the synchrotron emissions are precisely at the radio band. This is because the synchrotron emissions are likely self-absorbed at the radio band\footnote{The observed radio flux is probably the superimposed emissions of many outside optically thick emission regions \citep[e.g.,][]{1979ApJ...232...34B, 1985A&A...146..204G}.}. These lowest-energy electrons will IC scatter the soft seed photons and emit at the X-ray band. Therefore, careful SED modeling (especially modeling the IC X-ray spectra) is widely used to estimate the minimal electron energy and furthermore to calculate, for example, the jet power \citep[see, e.g.,][]{2008MNRAS.385..283C, 2010RAA....10..707C, 2014ApJS..215....5K, 2014Natur.515..376G}.

As discussed above, the X-ray spectra of LSP blazars and the rising part of the concave X-ray spectra of ISP blazars may be the character of the IC emission. For HSP blazars, soft X-rays are usually dominated by synchrotron emissions. The hard X-rays would be dominated by the low-energy IC emission part. However, due to the limited sensitivity of instruments in this energy range (above $\gtrsim$20 keV), we perhaps know the least about this lowest-energy part of the IC component.

Thanks to the hard X-ray energy range of the {\it Nuclear Spectroscopic Telescope Array} ({\it NuSTAR}; sensitive in 3-79 keV), the low-energy part of the IC component of some HSP blazars can be detected. For the first time, \citet{2016ApJ...827...55K} reported a clear detection of a hard X-ray excess (i.e., a concave X-ray spectrum) above $\gtrsim$20 keV of the HSP BL Lac object Mrk 421 during the low state \citep[see also PKS 2155-304;][]{2016ApJ...831..142M}. In this paper, we investigate the possibilities for the origin of this hard X-ray excess. In Section \ref{sec:excess}, the general properties of Mrk 421 are presented (especially the detection of the hard X-ray excess). Section \ref{sec:origin} provides a detailed discussion of the possible origins of this excess. Section \ref{sec:spinelayermodel} shows a spine/layer model, which can represent the hard X-ray excess. The discussion and conclusion are presented in Section \ref{sec:discussionconclusion}.


\section{The hard X-Ray excess of Mrk 421}
\label{sec:excess}

Mrk 421 is a typical HSP source, the first BL Lac object deteced by the Energetic Gamma Ray Experiment Telescope \citep[EGRET,][]{1992ApJ...401L..61L} at energy above 100 MeV and the first extragalactic source detected by imaging atmospheric Cherenkov telescopes (IACTs) at very high energy \citep[VHE; $\gtrsim$100 GeV,][]{1992Natur.358..477P}. It is so bright that it can be detected by modern IACTs within several minutes \citep{2015A&A...576A.126A}. Its X-ray and TeV emissions often show well-correlated variability, as do its optical and GeV emissions \citep[e.g.,][]{2016ApJS..222....6B, 2016ApJ...831..102K, 2016A&A...593A..91A, 2016A&A...591A..83S}. Including $\gamma$-rays, its broadband SED can be well fitted by a one-zone synchrotron self-Compton (SSC) model \citep[e.g.,][]{2013PASJ...65..109C, 2014MNRAS.442.3166Z, 2016MNRAS.463.4481Z, 2016ApJ...819..156B}. \citet{2000AIPC..515..165G} estimated the central black hole mass of order $\sim10^{9}$ $\rm M_{\odot}$ based on the imagery and broadline velocity dispersion. Its synchrotron component peaks at the hard X-ray band during the bright state and moves to the soft X-ray band when the source gets fainter \citep{2016ApJ...819..156B}. During the faintest state in MJD 56302, its synchrotron component peaks at too low a frequency to make it similar to ISP/LSP. Very recently, \citet{2016ApJ...827...55K} reported the detection of a significant excess above $\gtrsim$20 keV through {\it NuSTAR} observations at this faintest epoch \citep[see also the {\it BeppoSAX} observation; e.g.,][]{1999NuPhS..69..423F, 2000AIPC..510..313F}.

\section{The origin of the hard X-Ray excess}
\label{sec:origin}

\citet{2016ApJ...827...55K} compared the spectra of the hard X-ray excess with the simultaneous\footnote{The {\it Fermi}/LAT data are accumulated over roughly 1 week time intervals \citep[see][]{2016ApJ...819..156B}.} {\it Fermi}/LAT power-law $\gamma$-ray spectra \citep[the photon index $\Gamma=1.78\pm0.02$;][]{2016ApJ...819..156B}. They found that this excess follows the extension of the $\gamma$-ray spectra well; therefore, they argued that this hard X-ray excess must be the low-energy part of the SSC emissions of the same electron population that is producing the {\it Fermi}/LAT $\gamma$-ray. Actually, the broadband SED at this epoch had been well fitted by a one-zone SSC model \citep[up to the TeV band but except for the hard X-ray spectra;][]{2016ApJ...819..156B}. However, it can be seen that the observed hard X-ray excess cannot be reproduced by this one-zone modeling: the modeling underestimates the hard X-ray flux \citep[upper left panel of Figure 13 in][]{2016ApJ...819..156B}. The most likely possibility seems to be that a minimum electron energy that is too much larger ($\gamma_{\rm min}$) produces a ``cutoff" at the low energy part of the SSC component (i.e., very weak emissions at the hard X-ray band). This means that a smaller $\gamma_{\rm min}$ would increase the hard X-ray emissions and may represent the observed excess. The value of $\gamma_{\rm min}$ can be easily estimated when assuming that the hard X-ray emissions follow the power-law extension of the {\it Fermi}/LAT $\gamma$-ray spectra.

\begin{figure}
\begin{center}
{\includegraphics[width=1.0\linewidth]{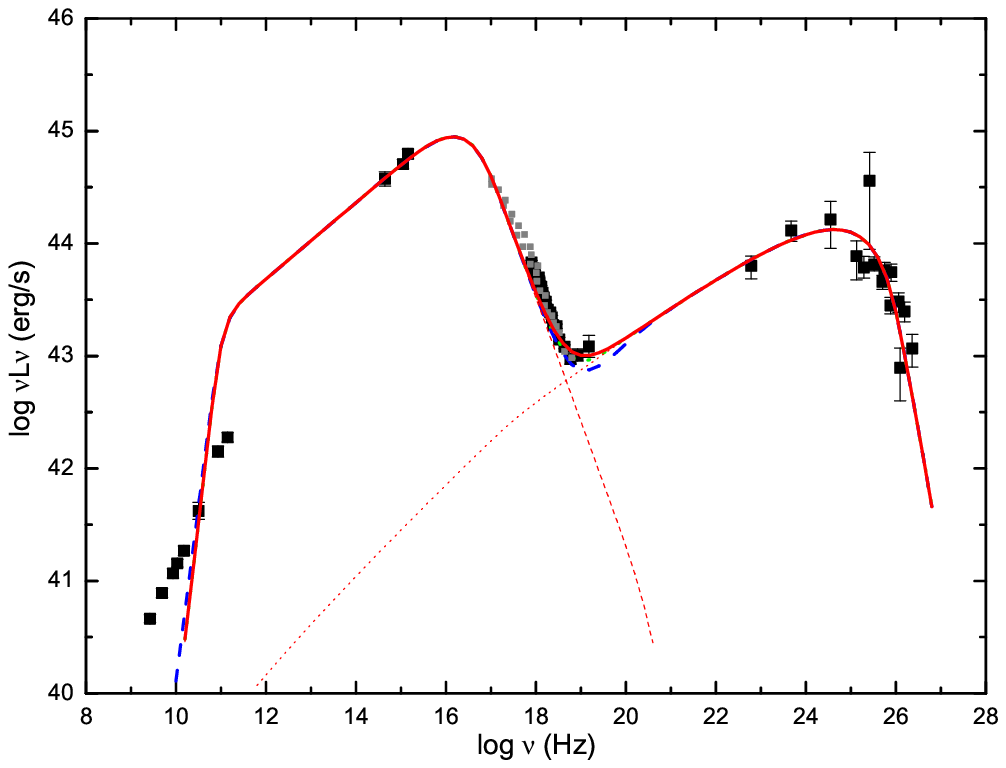}}
\end{center}
\caption{One-zone SSC model for SED modeling of Mrk 421 (synchrotron + SSC). The red solid line represents the total emissions with $\gamma_{\rm min}=2$, while the red dashed and dotted lines show the synchrotron and SSC emissions, respectively. The green dotted and blue dashed lines are for $\gamma_{\rm min}=20$ and 50, respectively. It can be seen that if one tries to fit the hard X-ray excess within the one-zone SSC model, the model will predict larger radio fluxes than the observed ones at 86.2 and 142.3 GHz.}
\label{sedonezone}
\end{figure}

We assume a one-zone model with a broken-power law electron energy distribution,
\begin{equation}
N(\gamma)=\left\{ \begin{array}{ll}
                    N_{0}\gamma ^{-p_1}  &  \mbox{ $\gamma_{\rm min}\leq \gamma \leq \gamma_{0}$} \\
            N_{0}\gamma _{\rm 0}^{p_2-p_1} \gamma ^{-p_2}  &  \mbox{ $\gamma _{\rm 0}<\gamma\leq\gamma_{\rm max}$.}
           \end{array}
       \right.
\label{Ngamma}
\end{equation}
In this model, the power-law distribution electrons below $\gamma_{0}$ produce the SED from the hard X-ray to the {\it Fermi}/LAT $\gamma$-ray emissions through the SSC process \citep[see][]{2016ApJ...819..156B}. Electrons around $\gamma_{\rm min}$ will IC synchrotron photons and produce the hard X-ray excess around $\gtrsim$20 KeV. We collect this simultaneous broadband SED from \citet{2016ApJ...819..156B} and \citet{2016ApJ...827...55K}, which are replotted in Figure \ref{sedonezone}. The synchrotron and SSC components peak at $\nu_{\rm sy}^{\rm p}\approx10^{16}$ and $\nu_{\rm SSC}^{\rm p}\approx10^{25}$ Hz, respectively, which correspond to emissions of electrons with energy around $\approx\gamma_{0}m_{\rm e}c^{2}$. As we know, the Klein-Nishina (KN) effect works at $\gamma_{0}h\nu_{\rm sy}/\delta\gtrsim m_{\rm e}c^{2}$ ($\delta$ is the Doppler beaming factor), and, in this case, the IC (SSC) photon energy approximates $\left(\sqrt{3}/2\right)h\nu_{\rm SSC}/\delta\approx\gamma_{0}m_{\rm e}c^{2}$. Combining these two formulas, we have $\delta\lesssim\sqrt{\left(h\nu_{\rm sy}\right)\left(h\nu_{\rm SSC}\right)}/ \left(m_{\rm e}c^{2}\right)$. Substituting the observed peak frequencies, we have $\delta\lesssim2.5$. This means that the KN effect can only work at peak frequency for a mild Doppler beaming factor (i.e., $\delta\lesssim2.5$), and we have $\gamma_{0}\approx7.0\times10^{4}/\delta$ in this case. Such a weak beaming effect seems unexpected for the HSP BL Lac object Mrk 421, because of that it often shows violent emissions and the derived $\delta$ from broadband SED modeling is much larger than this value \citep[e.g., $\delta=25$ for this epoch's SED modeling;][]{2016ApJ...819..156B}. For the case of IC scattering within the Thomsen regime, we have $\nu_{\rm SSC}\thickapprox\left(4/3\right)\gamma_{0}^{2}\nu_{\rm sy}$ and $\gamma_{0}h\nu_{\rm sy}/\delta\lesssim m_{\rm e}c^{2}$. Substituting the observed peak frequencies, we have $\delta\gtrsim2.5$ and $\gamma_{0}\approx2.7\times10^{4}$.

The IC frequency scales as $\nu_{\rm SSC}\thickapprox\left(4/3\right)\gamma^{2}\nu_{\rm sy}$. If $\gamma_{0}$ emits at $\nu_{\rm SSC}^{\rm p}\approx10^{25}$ Hz and $\gamma_{\rm min}$ emits at $\nu_{\rm SSC}^{\rm min}\approx4.8\times10^{18}$ Hz (i.e., the hard X-ray excess above $\approx20$ keV), one can estimate $\gamma_{\rm min}\approx\gamma_{0}\sqrt{\nu_{\rm SSC}^{\rm min}/\nu_{\rm SSC}^{\rm p}}\approx19$. Its synchrotron emissions could extend down to $\nu_{\rm sy}^{\rm min}\approx\nu_{\rm sy}^{\rm p}\gamma_{\rm min}^{2}/\gamma_{0}^{2}\approx4.8$ GHz. As disccused in Section \ref{sec:introduction}, the synchrotron emissions are usually synchrotron self-absorbed (SSA) at the radio band; therefore, the predicted radio flux (one-zone model) is usually below the observed one. Now, we will calculate this SSA radio flux (by $\gtrsim\gamma_{\rm min}$ electrons) and compare it with the observation.

Within the one-zone model, we can easily estimate the jet parameters when provided with a broadband SED (The emission region is assumed to be a homogeneous sphere with radius $R$ embedded in a magnetic field with strength $B$). For a SED at this epoch \citep[Figure \ref{sedonezone}; see also][]{2016ApJ...819..156B, 2016ApJ...827...55K}, the spectral indexes below and above the peak frequency are $\alpha_{1}=0.78\pm0.02$ and $\alpha_{2}=2.08\pm0.02$, respectively \citep[][]{2016ApJ...827...55K, 2011ApJ...736..131A, 2016ApJ...819..156B}. The peak luminosities of the synchrotron and SSC bumps are estimated to be $L_{\rm sy}^{\rm p}\approx1\times10^{45}$ and $L_{\rm SSC}^{\rm p}\approx2\times10^{44}$erg s$^{-1}$, respectively. Combined with the peak frequencies, one can estimate \citep[see e.g.,][]{1998ApJ...509..608T} the magnetic field strength $B\approx0.1$ Gs, the normalized electron density $N_{0}\approx5.4\times10^{4}$, the break energy $\gamma_{0}\approx2.7\times10^{4}$, the Doppler beaming factor $\delta=25$ \citep[assumed to be equal to the same value in][]{2016ApJ...819..156B}, the radius of the emission region $R\approx2.3\times10^{16}$cm, and the electron energy indexes below and above the break, $p_{1}=2.56$ and $p_{2}=5.16$, respectively.

The luminosity in the jet frame is given by \citep[see the Appendix for details; see also][]{1996ApJ...461..657B, 1999ApJ...514..138K},
\begin{equation}
L'(\nu')=2\pi^{2}R^{3}j(\nu')\frac{2\tau^{2}-1+\left(2\tau+1\right) e^{-2\tau}}{\tau^{3}},
\label{Lum}
\end{equation}
where the frequency $\nu'=(4/3)\gamma^{2}\nu_{\rm L}$, the emitting coefficient \citep[$\delta$-approximation; see the Appendix for details; see also][]{2002ApJ...575..667D, 2008ApJ...686..181F}
\begin{equation}
j(\nu')=\frac{\sigma_{\rm T}c}{8\pi}\frac{U_{\rm B}}{\nu_{\rm L}}\gamma N(\gamma),
\label{jem}
\end{equation}
and the optical depth
\begin{equation}
\tau=N_{0}R\left(\frac{3}{16}\right)^{2}\frac{4\sigma_{\rm T}c}{m_{\rm e}}\frac{U_{\rm B}}{\nu_{\rm L}^{3}}
\frac{1}{p_{1}+2}\gamma^{-p_{1}-4},
\label{tau}
\end{equation}
where $\nu_{\rm L}=eB/2\pi m_{\rm e}c$ is the Lamer frequency and the $U_{\rm B}=B^{2}/8\pi$ is the magnetic field energy density. The prime refers to the value in the jet frame. The frequency and luminosity in the active galactic nucleus (AGN) frame are transformed as $\nu=\delta\nu'$ and $\nu L(\nu)=\delta^{4}\nu' L'(\nu')$, respectively. The Pico Veleta telescope is used with the EMIR receiver to provide flux measurements at 86.2 GHz and 142.3 GHz \citep[see Figure \ref{sedonezone} and][]{2016ApJ...819..156B}. With the above jet parameters and Equation \ref{tau}, one can calculate the optical depth $\tau\approx3.3$ (corresponding electron energy $\gamma\approx80$) and $\tau\approx0.64$ ($\gamma\approx103$) at frequencies of 86.2 and 142.3 GHz, respectively. The expected luminosities (Equation \ref{Lum}) are $\nu L(\nu) \approx1.7\times10^{43}$ and $\approx5.6\times10^{43}$erg s$^{-1}$ at frequencies of 86.2 GHz and 142.3 GHz, respectively. It can be seen that both luminosities are contrary to the observed values, which are lower by $>$1 order of magnitude \citep[see Figure \ref{sedonezone} and ][]{2016ApJ...819..156B}. Even considering the error bar of the input parameters (i.e., the input peak luminosity and frequency and the spectral indexes), the expected luminosities at these two frequencies are larger than the observed ones. This means that even though the radio emission is SSA, the predicted radio fluxes (one-zone model) are still larger than the observed values, which implies that the electron power-law distribution cannot extend down to such a low energy (i.e., $\gamma_{\rm min}\approx19$). This indicates that the hard X-ray excess observed by {\it NuSTAR} at MJD 56302 cannot be produced by the low-energy part of the same electron population that produced the {\it Fermi}/LAT $\gamma$-rays.

A quantified one-zone SED modeling (synchrotron + SSC), including SSA and KN effects, is necessary to check the above estimation. For the model description, see the Appendix for detail \citep[see also][]{2012ApJ...748..119C, 2014ApJS..215....5K}. During our modeling, the Doppler beaming factor is assumed to be same as that in \citet{2016ApJ...819..156B}, i.e., $\delta=25$. For our purposes, the first step is setting the minimal electron energy $\gamma_{\rm min}=2$. The modeling SED curve is presented as the solid red line in Figure \ref{sedonezone}, and the other parameters are the magnetic field strength $B=0.114$Gs, the normalized electron density\footnote{It can be seen that the normalized electron density $N_{0}$ here is significantly smaller than the above estimation. This is due to the fact that the $N_{0}$ estimation depends significantly on the electron energy index below the break energy, $p_{1}=2.32$ here versus $p_{1}=2.56$ above, while the density at the break energy, $N(\gamma_{0})$, is similar for both.} $N_{0}=6.53\times10^{3}$, the break energy $\gamma_{0}=4.87\times10^{4}$, the radius of the emission region $R=2.29\times10^{16}$cm, the electron energy indexes below and above the break $p_{1}=2.32$ and $p_{2}=5.20$, and the maximal electron energy $\gamma_{\rm max}=100\gamma_{0}$. It can be seen that the model can reproduce the hard X-ray excess with this low value of $\gamma_{\rm min}$, while it predicts larger radio fluxes than ones observed at 86.2 and 142.3 GHz. The second step is that we increase the value of $\gamma_{\rm min}$ (keeping the other parameters unchanged) to see whether there is a $\gamma_{\rm min}$ value that can fit the hard X-ray excess and can also predict a lower radio flux. Two values, $\gamma_{\rm min}=20$ and 50, are set in the modeling, and results are shown in Figure \ref{sedonezone} as green dotted and blue dashed lines, respectively. It can be seen that even though the value increased to $\gamma_{\rm min}=50$, the predicted radio flux has almost no change due to the SSA effect, while the hard X-ray flux significantly decreases, which cannot represent the observed hard X-ray excess. These results confirm our above estimation that the hard X-ray excess observed by {\it NuSTAR} at MJD 56302 cannot be produced by the low-energy part of the same electron population that produced the {\it Fermi}/LAT $\gamma$-rays.

For the origin of the hard X-ray excess, another possibility is that this excess is actually the high-energy tail of synchrotron X-ray emissions. This requires a spectral pile-up in electron distribution at the highest energies. \citet{2016ApJ...827...55K} discussed this possibility and found that it is unlikely to work. The main reasons are as follows. (1) This high-energy pile-up bump could be achieved when the acceleration timescale equals the radiative-loss timescale at the limit for the perfect confinement of electrons within the emission zone \citep{2008ApJ...681.1725S}. However, this scenario predicts a flat power tail at lower electron energies, which is inconsistent with the X-ray observation of Mrk 421. (2) This high-energy tail can also be produced by the electron high-energy pile-up due to the reduction of the IC cross-section in the KN regime \citep{2005MNRAS.363..954M}. However, this requires IC cooling dominating over synchrotron cooling, which is not consistent with the fact that Mrk 421 is almost totally  dominated by synchrotron emission.

\section{The spine/layer jet}
\label{sec:spinelayermodel}

Here, we explore the possibility that the production of this hard X-ray excess is related to the spine/layer jet structure: a fast spine surrounded by a slower layer. The main reason for this choice is that Mrk 421 has been observed having a significant and clear spine/layer structure. In other aspects, the spine/layer structure can produce IC SEDs that are significantly different from those of the one-zone model \citep[see details below and, e.g.,][]{2005A&A...432..401G, 2016MNRAS.457.1352S}.

Mrk 421 shows a significant core-jet structure. Very long baseline interferometry (VLBI) observations can reveal a detailed jet structure at pc scale. Along the direction perpendicular to the jet axis, the fractional polarization shows significant double bumps increasing at the jet edges, which implies a spine/layer (also called spine-sheath) structure \citep{2010ApJ...723.1150P, 2014A&A...571A..54L}. This structure is supported by the similar distributions of the electron vector position angle (EVPA) and even the flux intensity \citep[the so-called limb brightening;][]{2010ApJ...723.1150P, 2014A&A...571A..54L}. The layer component usually moves slower than the spine. The relative opposite motion between the spine and layer will amplify the photon energy density from the spine (layer) in the frame of the layer (spine) and therefore produce a different (enhanced) IC emissions. This model has been widely and successfully used to explain the VHE $\gamma$-ray emissions from radio galaxies and blazars \citep[see, e.g.,][]{2005A&A...432..401G, 2008MNRAS.385L..98T, 2016MNRAS.457.1352S}.

The spine/layer model used here is same as that in \citet{2005A&A...432..401G}. For geometry, see Figure 1 in \citet{2005A&A...432..401G}. The layer is assumed to be a hollow cylinder with external radius $R_{2}$, internal radius $R$ and width $\Delta R_{\rm l}^{''}$ in the comoving frame of the layer\footnote{Double prime refers to values in the frame of the layer, and prime refers to values in the frame of the spine.}. For the cylindrical spine, the radius is $R$ and the width is $\Delta R_{\rm s}^{'}$. The spine and layer move with velocities $c\beta_{\rm s}$ and $c\beta_{\rm l}$, respectively, with $\Gamma_{\rm s}$ and $\Gamma_{\rm l}$ as the corresponding Lorentz factors. The relative velocity between the spine and the layer is then $\Gamma_{\rm rel}=\Gamma_{\rm s}\Gamma_{\rm l}(1-\beta_{\rm s}\beta_{\rm l})$. Following \citet{2005A&A...432..401G}, the radiation energy densities are considered as follows,
\begin{itemize}
  \item In the comoving frame of the layer, the radiation energy density $U_{\rm l}^{''}=L_{\rm l}^{''}/\left[\pi \left(R_{2}^{2}-R^{2}\right)c\right]$ \citep[slightly different from that of][to make sure that the radiation energy density is in the same format as that in the spine]{2005A&A...432..401G}. In the frame of the spine, this radiation energy density will be boosted to $U_{\rm l}^{'}=\Gamma_{\rm rel}^{2}U_{\rm l}^{''}$.
  \item In the comoving frame of the spine, the radiation energy within the spine is assumed to be $U_{\rm s}^{'}=L_{\rm s}^{'}/\left(\pi R^{2}c\right)$. The radiation energy density observed in the frame of the layer will be boosted by $\Gamma_{\rm rel}^{2}$ but also diluted (since the layer is larger than the spine) by a factor $\Delta R_{\rm s}^{''}/\Delta R_{\rm l}^{''}=\left(\Delta R_{\rm s}^{'}/\Gamma_{\rm rel}\right)/\Delta R_{\rm l}^{''}$.
\end{itemize}
The electron energy distributions in the spine and layer are all assumed to be broken power laws, as expressed in Equation \ref{Ngamma}, but with different parameters. In our modeling, we always assume $\Delta R_{\rm l}^{''}=30\Delta R_{\rm s}^{'}$, as in \citet{2005A&A...432..401G}. Figure \ref{sed} presents our SED modeling result, and the corresponding parameters are listed in Table \ref{jetparamters}. For the geometry parameters, $R=2.02\times10^{16}$, $R_{2}=2.42\times10^{16}$, and $\Delta R_{\rm s}^{'}=3.36\times10^{15}$cm. The red and blue lines represent the spine and layer emissions, respectively. While the dot-dashed lines are the synchrotron emissions, the dotted lines show the SSC emissions, and the dashed lines are for the IC emissions of seed photons originated externally from the layer/spine. It can been seen that, similar to one-zone modeling \citep{2016ApJ...819..156B}, the synchrotron + SSC emissions of the spine can produce almost a whole SED, except for the hard X-ray excess. This hard X-ray excess can be successfully represented by the process of seed photons (produced from the layer) being IC scattered by the nonthermal electrons within the spine.

\begin{table}
\tabletypesize{\tiny}
\setlength{\tabcolsep}{4.0pt}
\caption{The spine/Layer model parameters for Mrk 421.}
\begin{tabular}{@{}lcccccccccccc@{}}
\hline
\hline
     & $\theta$($^{\circ}$) &  $\Gamma$&  $\delta$ & $B$(Gs) &  $N_0$ & $\gamma_0$ & $\gamma_{\rm min}$& $\gamma_{\rm max}$ & $p_1$ & $p_2$\\
\hline
Spine & 2 & 21.4 & 27.5 & 0.23  & 996 & 29009 & 300 & 100$\gamma_{0}$& 2.0 & 5.2 \\
Layer & 2 & 4.7  & 9.1  & 0.0087& 593 & 326   & 2   & 100$\gamma_{0}$& 1.0 & 5.0 \\
\hline
\end{tabular}
\label{jetparamters}
\end{table}

\section{Discussion and Conclusions}
\label{sec:discussionconclusion}

The spine/layer jet model was initially studied by \citet{1989MNRAS.237..411S}. Such a scenario can explain some inconsistencies between the observations and the AGN standard unified model. Within the standard unified model, radio galaxies and blazars have the same relativistic jets but different viewing angles \citep{1995PASP..107..803U}. Because of the Doppler beaming factor significantly depending on the viewing angle, one can derive a number distribution as a function of viewing angle; roughly the number ratio between the radio galaxies and blazars. One finds that this number ratio (within the one-zone model) is inconsistent with observations \citep{2000A&A...358..104C}. This question can be solved if the jet is actually structured, e.g., a faster-moving spine surrounded by a slower-moving layer \citep{2000A&A...358..104C}. In the spine/layer model, jet emissions with small viewing angles will be dominated by the faster spine (for blazars), while the layer component will contribute more or even dominate the emissions in the case of a larger viewing angle (for radio galaxies). Therefore, in the case of the spine/layer model, one expects a larger number ratio between radio galaxies and blazars compared to the one-zone model \citep{2000A&A...358..104C}. As discussed above, the relative opposite motion between the spine and the layer will amplify the photon energy density produced from the spine (layer) in the frame of the layer (spine) by $\sim\Gamma_{\rm rel}^{2}$ (see Section \ref{sec:spinelayermodel}). Therefore, one expects that the IC emissions will be enhanced relative to the one-zone model \citep{2005A&A...432..401G}. This spine/layer model has been widely and successfully used to explain the VHE $\gamma$-ray emissions of radio galaxies and blazars \citep[see, e.g., ][]{2005A&A...432..401G, 2008MNRAS.385L..98T, 2016MNRAS.457.1352S}. The orphan flare is an interesting phenomenon in blazar observations. Recently, \citet{2015ApJ...804..111M} proposed that, when a faster-moving spine (emission zone) moves across a slower-moving (or steady) layer/ring, the sudden increase of the energy density of external seed photons (produced from the layer) will be IC scattered by the spine electrons and produce an orphan flare in the $\gamma$-ray band \citep[see also][]{2016arXiv161109953M}. The origin of the spine/layer structure is, however, unclear. It could arise directly from a jet launching process in which the external layer is ejected from the accretion disk while the central spine is fueled from the black hole ergosphere, as shown in some numerical simulations \citep[e.g.,][]{2007ApJ...662..835M, 2010A&A...521A..67M}. Alternatively, the Kelvin-Helmholtz instability can occur when there is velocity shear in a single continuous fluid, which can create the spine/layer structure seen in 3C 273 \citep[][]{2001Sci...294..128L}. The third possibility is that the falling of the magnetized accretion flow leads to accumulation of the inflating toroidal magnetic field inside the accretion disk, which can produce a magnetic tower jet. The magnetic tower jet presents a central helical magnetic field ``spine" surrounded by a reversed magnetic field ``sheath," which could be a prototype of a spine/layer structure \citep[see, e.g.,][]{2004ApJ...605..307K, 2006ApJ...643...92L}.

For the first time, \citet{2016ApJ...827...55K} reported the detection of the hard X-ray excess above $\gtrsim$20 keV in the HSP BL Lac object Mrk 421. This feature offers important information for our understanding of jet physics. Many probabilities may account for this excess. However, as discussed above, it cannot be the high-energy tail of the synchrotron X-ray emissions, because the requirement of high-energy pile-up in electron energy distribution cannot be achieved due to the theoretical predictions being in conflict with the observations \citep[the spectral index/Compton dominance; see Section \ref{sec:origin} and][]{2016ApJ...827...55K}. Alternatively, it seems to be the low-energy tail of the {\it Fermi}/LAT $\gamma$-ray spectra. As discussed in Section \ref{sec:origin}, this possibility requires the minimal electron energy down to $\gamma_{\rm min}\approx19$. Such a lower electron energy predicts a very strong radio emission, even with the SSA effect, which is larger than the observed radio flux. Therefore, the hard X-ray excess cannot be the low-energy tail of the {\it Fermi}/LAT $\gamma$-ray spectra.

\begin{figure}
\begin{center}
{\includegraphics[width=1.0\linewidth]{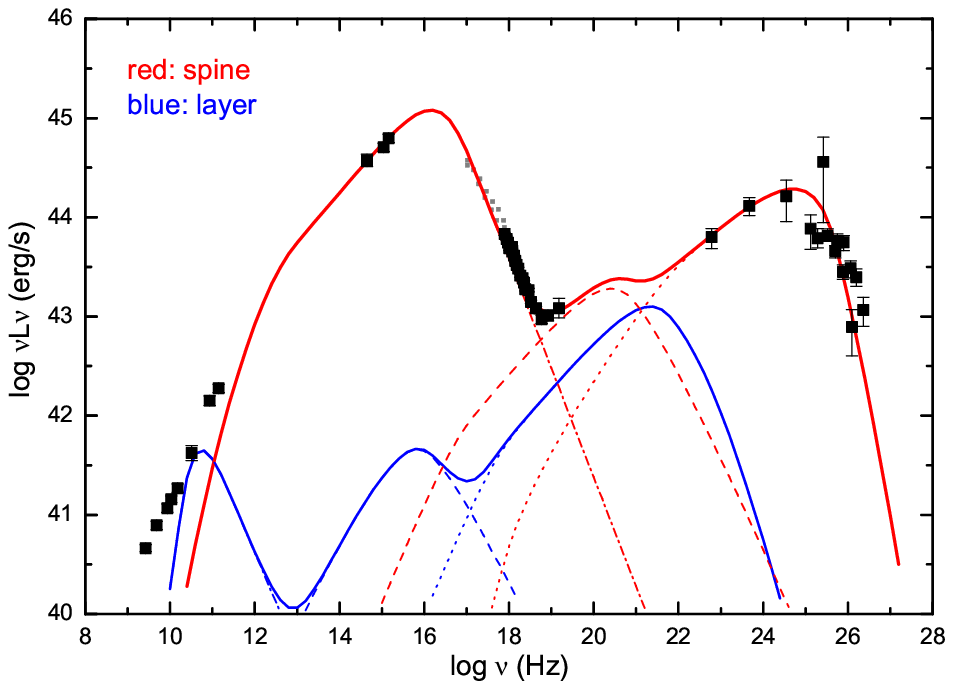}}
\end{center}
\caption{Spine/layer model for SED modeling of Mrk 421. The red and blue lines represent the spine and layer emissions, respectively. While the dot-dashed lines are the synchrotron emissions, the dotted lines show the SSC emissions, and the dashed lines are for the IC emissions of seed photons originated externally from the layer/spine.}
\label{sed}
\end{figure}

Because of that, Mrk 421 shows a clear spine/layer structure from VLBI observations (e.g., the limb brightening, the double-hump distribution of fractional polarization, and the EPVA along the perpendicular jet direction). We investigate the possibilities of the hard X-ray excess produced related to this structure. We find that the spine emissions provide good-modeling of the broadband SED except for the hard X-ray excess, which is similar to one-zone modeling \citep[see Figure \ref{sed} and][]{2016ApJ...819..156B}. The hard X-ray excess can be well represented by the synchrotron photons from the layer being IC scattered by the spine electrons (see Figure \ref{sed}). Until now, about 49 HSP BL Lacs\footnote{http://tevcat.uchicago.edu/} have been detected that have VHE emissions. The broadband SEDs of some of them can be well fitted by the one-zone model \citep[e.e.,][]{2009A&A...504..821P, 2012ApJ...752..157Z, 2014Natur.515..376G, 2014MNRAS.439.2933Y, 2017MNRAS.464..599D}. However, it should be noted that these well-modeled SEDs do not include the rising part at the hard X-ray band. If these sources have hard X-ray spectra similar to that of Mrk 421 (i.e., the hard X-ray excess/the concave feature), it can be seen that almost all of these SED modelings underestimate their hard X-ray emissions. Actually, some LSP BL Lacs have a similar question. The LSP BL Lacs present very hard X-ray spectra, which may be the rising part of the second IC component. However, the one-zone SSC model is also much harder to fit to their broadband SEDs \citep{2014MNRAS.439.2933Y}, which are usually modeled by a two-zone model or a one-zone SSC + EC (external Comptoton) model \citep[e.g.,][]{2011MNRAS.414.2674G, 2011ApJ...730..101A}. For our SED modeling of Mrk 421, it should be noted that the model used is just a simple toy model, i.e., using two distinct regions (spine and layer) instead of continuing the distribution from jet axis to edge, although our modeled parameters are among the typical values of blazars \citep[e.g.,][]{2010MNRAS.402..497G, 2014Natur.515..376G, 2012ApJ...752..157Z}.

The HSP BL Lac PKS 2155-304 has also been seen to present a flattened spectrum above the $\gtrsim$4 keV during the low state \citep[observations by {\it BeppoSAX}, {\it XMM-Newton}, and {\it NuSTAR};][]{1998A&A...333L...5G, 1999ApJ...521..552C, 2008ApJ...682..789Z, 2016ApJ...831..142M}. This flattened X-ray spectrum had been considered to be SSC emission of the lowest-energy electrons, and the broadband SED was fitted within the one-zone SSC model by \citet{2016ApJ...831..142M}. The minimal electron energy down to $\gamma_{\rm min}\approx1$ is required to reproduce the hard X-ray feature. Because of that, the jet power significantly depends on the minimal electron energy. Such a small $\gamma_{\rm min}$ requires a very large jet power (assuming one cold proton per electron), $\sim1.2\times10^{47}$erg s$^{-1}$, which requires the accretion rate to exceed the Eddington rate, even assuming high efficiency of conversion of the accretion power to jet power \citep{2016ApJ...831..142M}. As suggested by \citet{2016MNRAS.457.1352S}, the spine/layer model is less demanding of jet power than the one-zone model and can reproduce the basic features of $\gamma$-ray events. In another aspect, the radio data to compare with the SED modeling of PKS 2155-304 are lacking \citep[see Figure 4 in][]{2016ApJ...831..142M}.

The jet power of Mrk 421 (including both spine and layer) can be easily calculated when provided with these SED modeling parameters (see Table \ref{jetparamters}): $P_{\rm jet}\approx2.0\times10^{45}$erg s$^{-1}$, which is $\sim55$ times the jet radiation power (assuming $\Gamma=\delta$), and $P_{\rm rad}\approx L_{\rm bol}/\Gamma^{2}\approx3.6\times10^{43}$erg s$^{-1}$, where $L_{\rm bol}$ is the bolometric jet luminosity.  Therefore, the jet carries $\sim$1.6\% of the Eddington luminosity, which is consistent with the fact that HSP BL Lacs (and Mrk 421 in particular) accrete via inefficient, low-accretion-rate, or advection-dominated accretion flow \citep[for a recent overview, see][]{2014ARA&A..52..529Y}.

\acknowledgments

We thank the anonymous referee for insightful comments and constructive suggestions. This work is supported by the NSFC (grants 11233006 and U1431123) and the Key Laboratory for the Structure and Evolution of Celestial Objects, Chinese Academy of Sciences (OP20140X)

\appendix
\label{append}

\section{One-zone SSC model}

The one-zone SSC model adopted in this paper assumes a homogeneous and isotropic emission region, which is a sphere with radius $R$ that has a uniform magnetic field with strength $B$ and a uniform electron energy distribution as described by Equation \ref{Ngamma}. The emission region moves relativistically with a Lorentz factor $\Gamma=1/\sqrt{1-\beta^{2}}$ and a viewing angle $\theta$, which forms the Doppler beaming factor $\delta=1/\left[\Gamma\left(1-\beta\cos\theta\right)\right]$. The frequency and luminosity transform from the jet to AGN frames as $\nu=\delta\nu'$ and $\nu L(\nu)=\delta^{4}\nu' L'(\nu')$, respectively.

\begin{figure}
\begin{center}
{\includegraphics[width=1.0\linewidth]{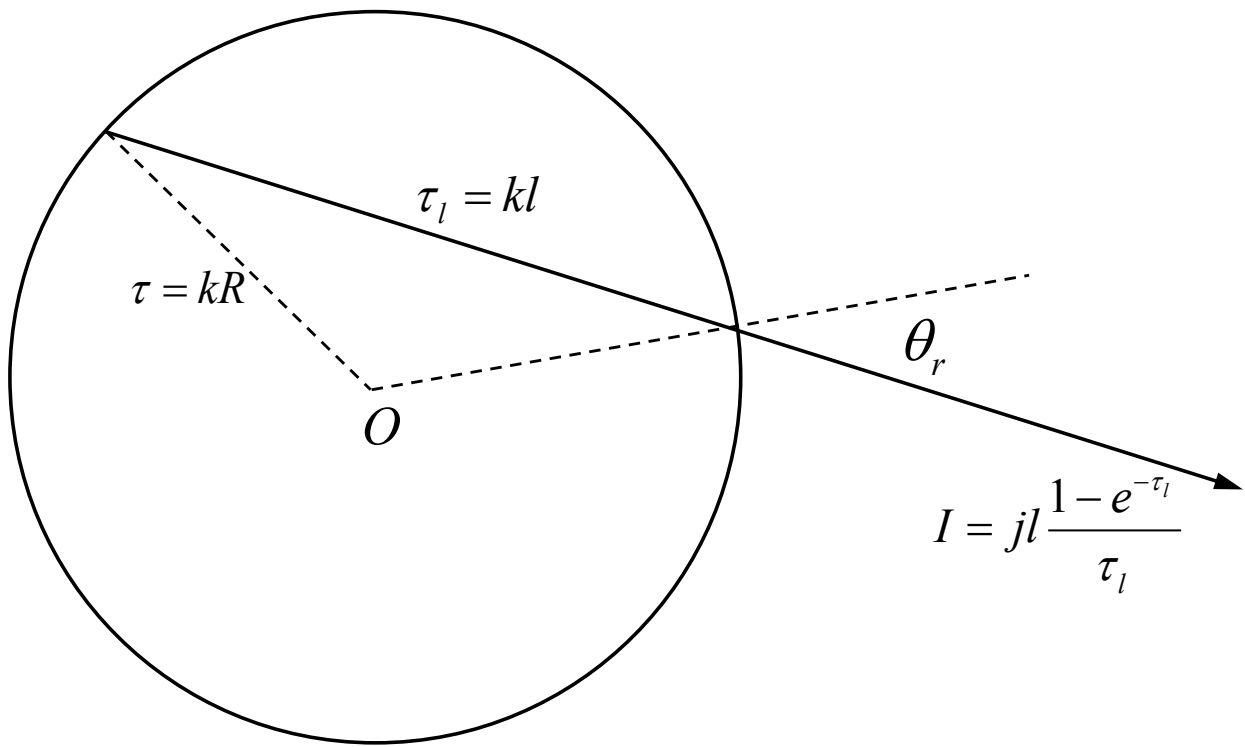}}
\end{center}
\caption{Sketch of the emission region of a homogeneous and isotropic sphere. Note that the intensity ($I$) is angle ($\theta_{r}$) dependent.}
\label{figA_tau}
\end{figure}

The Figure \ref{figA_tau} shows a sketch of the emission region. The intensity at the sphere surface can be easily derived,
\begin{equation}
I(\nu',\theta_{r})=j(\nu')l\frac{1-e^{-\tau_{l}}}{\tau_{l}},
\label{eqA_intensity}
\end{equation}
where $j(\nu')$ is the emitting coefficient, $\tau_{l}=k(\nu')l$ is the optical depth, and $k(\nu')$ is the absorption coefficient. Note that the intensity is angle ($\theta_{r}$) dependent. The total luminosity can be derived through integrating the angle-dependent intensity \citep[see, e.g.,][]{1996ApJ...461..657B, 1999ApJ...514..138K},
\begin{equation}
L'(\nu')=2\pi^{2}R^{3}j(\nu')\frac{2\tau^{2}-1+\left(2\tau+1\right) e^{-2\tau}}{\tau^{3}},
\label{eqA_luminosity}
\end{equation}
where $\tau=k(\nu')R$. At the limit of small $\tau$, one has $L'(\nu')\simeq\frac{4}{3}\pi R^{3}\cdot4\pi j(\nu')\cdot\left(1-\frac{3}{4}\tau\right)$.

The emitting coefficient of synchrotron emission is given by \citep{1970RvMP...42..237B, 1979rpa..book.....R}
\begin{equation}
j_{sy}(\nu')=\frac{1}{4\pi}\int N(\gamma)P_{sy}(\nu',\gamma)d\gamma,
\label{eqA_jsy}
\end{equation}
and the absorption coefficient,
\begin{equation}
k_{sy}(\nu')=\frac{-1}{8\pi m_{e}\nu'^{2}}\int\gamma^{2}P_{sy}(\nu',\gamma)\frac{d}{d\gamma}\left(\frac{N(\gamma)}{\gamma^{2}}\right)d\gamma,
\label{eqA_ksy}
\end{equation}
where the synchrotron emission power of a single electron is
\begin{equation}
P_{sy}(\nu',\gamma)=\frac{2\pi\sqrt{3}e^{2}\nu_{L}}{c}\frac{\nu'}{(3/2)\gamma^{2}\nu_{L}}\int_{\frac{\nu'}{(3/2)\gamma^{2}\nu_{L}}}^{+\infty}K_{5/3}(t)dt
\label{eqA_Psy}
\end{equation}
where $K_{5/3}(t)$ is the modified Bessel function of order 5/3 and $\nu_{\rm L}=eB/2\pi m_{\rm e}c$ is the Lamer frequency.

For SSC, the seed photon energy density (synchrotron emission itself) at the center of the emission sphere may be different from that at the edge of the emission sphere due to light transporting. At a distance $r$ from the center, the energy density is given by $u_{sy}(\nu',r)=\frac{1}{c}\oint I_{sy}(\nu',r,\theta_{r})d\Omega$. Because of that the seed photons are dominated by optically thin emissions, one has an analytic solution of the photon energy density (assuming $\tau=0$)
\begin{equation}
u_{sy}(\nu',r)=\frac{3}{4}\frac{L_{sy}'(\nu')}{4\pi R^{2}c}\left(2+\frac{1-r_{\ast}^{2}}{r_{\ast}}\ln\frac{1+r_{\ast}}{1-r_{\ast}}\right),
\label{eqA_u}
\end{equation}
where $r_{\ast}=r/R$. At the edge of the emission sphere, one has $u_{sy}(\nu',R)=(3/2)L_{sy}'(\nu')/(4\pi R^{2}c)$, and, at the center, $u_{sy}(\nu',0)=3L_{sy}'(\nu')/(4\pi R^{2}c)$. The average value, $\overline{u}_{sy}(\nu')=(9/4)L_{sy}'(\nu')/(4\pi R^{2}c)$, is used in our calculation. In this case, the emitting coefficient of SSC emission is \citep{1970RvMP...42..237B, 1979rpa..book.....R}
\begin{equation}
j_{ssc}(\nu')=\frac{1}{4\pi}\int N(\gamma)P_{ssc}(\nu',\gamma)d\gamma,
\label{eqA_jssc}
\end{equation}
where the SSC emission power of a single electron is
\begin{equation}
P_{ssc}(\nu',\gamma)=8\pi r_{0}^{2}ch\int f(\gamma,\nu_{i},\nu')n_{ph}(\nu_{i})d\nu_{i},
\label{eqA_Pssc}
\end{equation}
where $r_{0}=e^{2}/\left(m_{e}c^{2}\right)$ and $n_{ph}(\nu_{i})=\overline{u}_{sy}(\nu_{i})/h\nu_{i}$ is the seed photon number density. The function $f(\gamma,\nu_{i},\nu')$ is given by \citep{1970RvMP...42..237B}
\begin{equation}
f(\gamma,\nu_{i},\nu')=\left\{ \begin{array}{ll}
                    x\left[2q\ln q+1+q-2q^{2}+\frac{1}{2}\frac{(\Sigma q)^{2}}{1+\Sigma q}(1-q)\right] \mbox{ $0\leq q \leq 1$}\\
            0   \mbox{ $else$}
           \end{array}
       \right.
\label{eqA_fssc}
\end{equation}
where $x=\nu'/(4\gamma^{2}\nu_{i})$, $q=E/[\Sigma(1-E)]$, $E=h\nu'/(\gamma m_{e}c^{2})$, and $\Sigma=4\gamma h\nu_{i}/(m_{e}c^{2})$.

Note that, because the synchrotron emission of a single electron is not very broad in the frequency space \citep[Equation \ref{eqA_Psy}; see also][]{1970RvMP...42..237B, 1979rpa..book.....R}, one sometimes assumes that all emissions focus on a particular frequency, i.e., the $\delta$-approximation (monochromatic approximation). Using the following equation instead of Equation \ref{eqA_Psy},
\begin{equation}
P_{sy}(\nu',\gamma)\approx\frac{4}{3}\sigma_{T}cU_{B}\gamma^{2}\delta(\nu'-\frac{4}{3}\nu_{L}\gamma^{2}).
\label{eqA_Psy_app}
\end{equation}
In this case, the emitting coefficient (Equation \ref{eqA_jsy}) and the absorption coefficient (Equation \ref{eqA_ksy}) will be reduced to Equations \ref{jem} and \ref{tau} \citep[the optical depth $\tau=k(\nu')R$; see also][]{2002ApJ...575..667D, 2008ApJ...686..181F}.

\newpage



\begin{thebibliography}{99}

\bibitem[Abdo et al.(2010)]{2010ApJ...715..429A} Abdo, A.~A., Ackermann, M., Ajello, M., et al.\ 2010, \apj, 715, 429


\bibitem[Abdo et al.(2011)]{2011ApJ...730..101A} Abdo, A.~A., Ackermann, M., Ajello, M., et al.\ 2011, \apj, 730, 101


\bibitem[Abdo et al.(2011)]{2011ApJ...736..131A} Abdo, A.~A., Ackermann, M., Ajello, M., et al.\ 2011, \apj, 736, 131


\bibitem[Aharonian(2000)]{2000NewA....5..377A} Aharonian, F.~A.\ 2000, \na, 5, 377


\bibitem[Ahnen et al.(2016)]{2016A&A...593A..91A} Ahnen, M.~L., Ansoldi, S., Antonelli, L.~A., et al.\ 2016, \aap, 593, A91


\bibitem[Aleksi{\'c} et al.(2015)]{2015A&A...576A.126A} Aleksi{\'c}, J., Ansoldi, S., Antonelli, L.~A., et al.\ 2015, \aap, 576, A126


\bibitem[Atoyan \& Dermer(2003)]{2003ApJ...586...79A} Atoyan, A.~M., \& Dermer, C.~D.\ 2003, \apj, 586, 79


\bibitem[Balokovi{\'c} et al.(2016)]{2016ApJ...819..156B} Balokovi{\'c}, M., Paneque, D., Madejski, G., et al.\ 2016, \apj, 819, 156


\bibitem[Bartoli et al.(2016)]{2016ApJS..222....6B} Bartoli, B., Bernardini, P., Bi, X.~J., et al.\ 2016, \apjs, 222, 6


\bibitem[Blandford \& K{\"o}nigl(1979)]{1979ApJ...232...34B} Blandford, R.~D., \& K{\"o}nigl, A.\ 1979, \apj, 232, 34


\bibitem[Blandford \& Rees(1978)]{1978bllo.conf..328B} Blandford, R.~D., \& Rees, M.~J.\ 1978, BL Lac Objects, 328


\bibitem[Bloom \& Marscher(1996)]{1996ApJ...461..657B} Bloom, S.~D., \& Marscher, A.~P.\ 1996, \apj, 461, 657


\bibitem[Blumenthal \& Gould(1970)]{1970RvMP...42..237B} Blumenthal, G.~R., \& Gould, R.~J.\ 1970, Reviews of Modern Physics, 42, 237


\bibitem[Cao \& Wang(2013)]{2013PASJ...65..109C} Cao, G., \& Wang, J.\ 2013, \pasj, 65, 109


\bibitem[Celotti \& Ghisellini(2008)]{2008MNRAS.385..283C} Celotti, A., \& Ghisellini, G.\ 2008, \mnras, 385, 283


\bibitem[Chen et al.(2010)]{2010RAA....10..707C} Chen, L., Bai, J.-M., Zhang, J., \& Liu, H.-T.\ 2010, Research in Astronomy and Astrophysics, 10, 707


\bibitem[Chen et al.(2012)]{2012ApJ...748..119C} Chen, L., Cao, X., \& Bai, J.~M.\ 2012, \apj, 748, 119


\bibitem[Chiaberge et al.(2000)]{2000A&A...358..104C} Chiaberge, M., Celotti, A., Capetti, A., \& Ghisellini, G.\ 2000, \aap, 358, 104


\bibitem[Chiappetti et al.(1999)]{1999ApJ...521..552C} Chiappetti, L., Maraschi, L., Tavecchio, F., et al.\ 1999, \apj, 521, 552


\bibitem[Dermer \& Schlickeiser(2002)]{2002ApJ...575..667D} Dermer, C.~D., \& Schlickeiser, R.\ 2002, \apj, 575, 667


\bibitem[Dermer \& Schlickeiser(1993)]{1993ApJ...416..458D} Dermer, C.~D., \& Schlickeiser, R.\ 1993, \apj, 416, 458


\bibitem[Ding et al.(2017)]{2017MNRAS.464..599D} Ding, N., Zhang, X., Xiong, D.~R., \& Zhang, H.~J.\ 2017, \mnras, 464, 599


\bibitem[Finke et al.(2008)]{2008ApJ...686..181F} Finke, J.~D., Dermer, C.~D., \& B{\"o}ttcher, M.\ 2008, \apj, 686, 181-194


\bibitem[Fossati et al.(2000)]{2000AIPC..510..313F} Fossati, G., Celotti, A., Chiaberge, M., \& Zhang, Y.~H.\ 2000, American Institute of Physics Conference Series, 510, 313


\bibitem[Fossati et al.(1999)]{1999NuPhS..69..423F} Fossati, G., Chiappetti, L., Celotti, A., et al.\ 1999, Nuclear Physics B Proceedings Supplements, 69, 423


\bibitem[Ghisellini et al.(1985)]{1985A&A...146..204G} Ghisellini, G., Maraschi, L., \& Treves, A.\ 1985, \aap, 146, 204


\bibitem[Ghisellini \& Tavecchio(2009)]{2009MNRAS.397..985G} Ghisellini, G., \& Tavecchio, F.\ 2009, \mnras, 397, 985


\bibitem[Ghisellini et al.(2005)]{2005A&A...432..401G} Ghisellini, G., Tavecchio, F., \& Chiaberge, M.\ 2005, \aap, 432, 401


\bibitem[Ghisellini et al.(2011)]{2011MNRAS.414.2674G} Ghisellini, G., Tavecchio, F., Foschini, L., \& Ghirlanda, G.\ 2011, \mnras, 414, 2674


\bibitem[Ghisellini et al.(2010)]{2010MNRAS.402..497G} Ghisellini, G., Tavecchio, F., Foschini, L., et al.\ 2010, \mnras, 402, 497


\bibitem[Ghisellini et al.(2014)]{2014Natur.515..376G} Ghisellini, G., Tavecchio, F., Maraschi, L., Celotti, A., \& Sbarrato, T.\ 2014, \nat, 515, 376


\bibitem[Giommi et al.(1998)]{1998A&A...333L...5G} Giommi, P., Fiore, F., Guainazzi, M., et al.\ 1998, \aap, 333, L5


\bibitem[Gorham et al.(2000)]{2000AIPC..515..165G} Gorham, P.~W., van Zee, L., Unwin, S.~C., \& Jacobs, C.~S.\ 2000, American Institute of Physics Conference Series, 515, 165


\bibitem[Kang et al.(2014)]{2014ApJS..215....5K} Kang, S.-J., Chen, L., \& Wu, Q.\ 2014, \apjs, 215, 5


\bibitem[Kapanadze et al.(2016)]{2016ApJ...831..102K} Kapanadze, B., Dorner, D., Vercellone, S., et al.\ 2016, \apj, 831, 102


\bibitem[Kataoka et al.(1999)]{1999ApJ...514..138K} Kataoka, J., Mattox, J.~R., Quinn, J., et al.\ 1999, \apj, 514, 138


\bibitem[Kataoka \& Stawarz(2016)]{2016ApJ...827...55K} Kataoka, J., \& Stawarz, {\L}.\ 2016, \apj, 827, 55


\bibitem[Kato et al.(2004)]{2004ApJ...605..307K} Kato, Y., Mineshige, S., \& Shibata, K.\ 2004, \apj, 605, 307


\bibitem[Li et al.(2006)]{2006ApJ...643...92L} Li, H., Lapenta, G., Finn, J.~M., Li, S., \& Colgate, S.~A.\ 2006, \apj, 643, 92


\bibitem[Lico et al.(2014)]{2014A&A...571A..54L} Lico, R., Giroletti, M., Orienti, M., et al.\ 2014, \aap, 571, A54


\bibitem[Lin et al.(1992)]{1992ApJ...401L..61L} Lin, Y.~C., Bertsch, D.~L., Chiang, J., et al.\ 1992, \apjl, 401, L61


\bibitem[Lobanov \& Zensus(2001)]{2001Sci...294..128L} Lobanov, A.~P., \& Zensus, J.~A.\ 2001, Science, 294, 128


\bibitem[M{\"u}cke et al.(2003)]{2003APh....18..593M} M{\"u}cke, A., Protheroe, R.~J., Engel, R., Rachen, J.~P., \& Stanev, T.\ 2003, Astroparticle Physics, 18, 593


\bibitem[MacDonald et al.(2016)]{2016arXiv161109953M} MacDonald, N.~R., Jorstad, S.~G., \& Marscher, A.~P.\ 2016, arXiv:1611.09953


\bibitem[MacDonald et al.(2015)]{2015ApJ...804..111M} MacDonald, N.~R., Marscher, A.~P., Jorstad, S.~G., \& Joshi, M.\ 2015, \apj, 804, 111


\bibitem[Madejski et al.(2016)]{2016ApJ...831..142M} Madejski, G.~M., Nalewajko, K., Madsen, K.~K., et al.\ 2016, \apj, 831, 142


\bibitem[Mannheim(1993)]{1993A&A...269...67M} Mannheim, K.\ 1993, \aap, 269, 67


\bibitem[Maraschi et al.(1992)]{1992ApJ...397L...5M} Maraschi, L., Ghisellini, G., \& Celotti, A.\ 1992, \apjl, 397, L5


\bibitem[Meliani et al.(2010)]{2010A&A...521A..67M} Meliani, Z., Sauty, C., Tsinganos, K., Trussoni, E., \& Cayatte, V.\ 2010, \aap, 521, A67


\bibitem[Mizuno et al.(2007)]{2007ApJ...662..835M} Mizuno, Y., Hardee, P., \& Nishikawa, K.-I.\ 2007, \apj, 662, 835


\bibitem[Moderski et al.(2005)]{2005MNRAS.363..954M} Moderski, R., Sikora, M., Coppi, P.~S., \& Aharonian, F.\ 2005, \mnras, 363, 954


\bibitem[Paggi et al.(2009)]{2009A&A...504..821P} Paggi, A., Massaro, F., Vittorini, V., et al.\ 2009, \aap, 504, 821


\bibitem[Piner et al.(2010)]{2010ApJ...723.1150P} Piner, B.~G., Pant, N., \& Edwards, P.~G.\ 2010, \apj, 723, 1150


\bibitem[Punch et al.(1992)]{1992Natur.358..477P} Punch, M., Akerlof, C.~W., Cawley, M.~F., et al.\ 1992, \nat, 358, 477


\bibitem[Rybicki \& Lightman(1979)]{1979rpa..book.....R} Rybicki, G.~B., \& Lightman, A.~P.\ 1979, New York, Wiley-Interscience, 1979.~393 p.,


\bibitem[Sikora et al.(1994)]{1994ApJ...421..153S} Sikora, M., Begelman, M.~C., \& Rees, M.~J.\ 1994, \apj, 421, 153


\bibitem[Sikora et al.(2016)]{2016MNRAS.457.1352S} Sikora, M., Rutkowski, M., \& Begelman, M.~C.\ 2016, \mnras, 457, 1352


\bibitem[Sinha et al.(2016)]{2016A&A...591A..83S} Sinha, A., Shukla, A., Saha, L., et al.\ 2016, \aap, 591, A83


\bibitem[Sol et al.(1989)]{1989MNRAS.237..411S} Sol, H., Pelletier, G., \& Asseo, E.\ 1989, \mnras, 237, 411


\bibitem[Stawarz \& Petrosian(2008)]{2008ApJ...681.1725S} Stawarz, {\L}., \& Petrosian, V.\ 2008, \apj, 681, 1725-1744


\bibitem[Tavecchio \& Ghisellini(2008)]{2008MNRAS.385L..98T} Tavecchio, F., \& Ghisellini, G.\ 2008, \mnras, 385, L98


\bibitem[Tavecchio et al.(1998)]{1998ApJ...509..608T} Tavecchio, F., Maraschi, L., \& Ghisellini, G.\ 1998, \apj, 509, 608


\bibitem[Urry \& Padovani(1995)]{1995PASP..107..803U} Urry, C.~M., \& Padovani, P.\ 1995, \pasp, 107, 803


\bibitem[Yan et al.(2014)]{2014MNRAS.439.2933Y} Yan, D., Zeng, H., \& Zhang, L.\ 2014, \mnras, 439, 2933


\bibitem[Yuan \& Narayan(2014)]{2014ARA&A..52..529Y} Yuan, F., \& Narayan, R.\ 2014, \araa, 52, 529


\bibitem[Zhang et al.(2012)]{2012ApJ...752..157Z} Zhang, J., Liang, E.-W., Zhang, S.-N., \& Bai, J.~M.\ 2012, \apj, 752, 157


\bibitem[Zhang(2008)]{2008ApJ...682..789Z} Zhang, Y.~H.\ 2008, \apj, 682, 789-797


\bibitem[Zheng et al.(2014)]{2014MNRAS.442.3166Z} Zheng, Y.~G., Kang, S.~J., \& Li, J.\ 2014, \mnras, 442, 3166


\bibitem[Zhu et al.(2016)]{2016MNRAS.463.4481Z} Zhu, Q., Yan, D., Zhang, P., et al.\ 2016, \mnras, 463, 4481



\end{thebibliography}
\end{document}